\documentclass[useAMS,usenatbib]{mn2e}

\usepackage{times}
\usepackage{amsmath}
\usepackage{amssymb}
\usepackage{graphicx}
\usepackage{graphics}

\title[OII Emission from Cluster Galaxies]
{The Evolution of [OII] Emission from Cluster Galaxies}
\author[Nakata et al.]{
Fumiaki Nakata$^{1}$,\thanks{E-mail: fumiaki.nakata@durham.ac.uk}
Richard G. Bower$^{1}$, Michael L. Balogh$^{2}$, David J. Wilman$^{1,3}$ \\
$^{1}$Department of Physics, University of Durham, South Road, Durham DH1
3LE, UK \\
$^{2}$Department of Physics, University of Waterloo, Waterloo, Ontario
N2L 3G1, Canada \\
$^{3}$Max-Planck Institut f\"{u}r extraterrestrische Physik,
Giessenbachstrasse, D-85748 Garching, Germany \\
}

\begin{document}

\date{Accepted ???. Received ???; in original form ???}

\pagerange{\pageref{firstpage}--\pageref{lastpage}} \pubyear{2004}

\maketitle

\label{firstpage}

\begin{abstract}

We investigate the evolution of the star formation rate in cluster
galaxies. We complement data from the CNOC1 cluster survey
($0.15<z<0.6$) with measurements from galaxy clusters in the 2dF
galaxy redshift survey ($0.05<z<0.1$) and measurements from recently
published work on higher redshift clusters, up to almost $z=1$. We
focus our attention on galaxies
in the cluster core, ie.\ galaxies with $r<0.7h^{-1}_{70}$~Mpc.
Averaging over clusters in redshift bins, we find that the 
fraction of galaxies with strong [OII] emission is
$\lesssim 20$\% in cluster cores, and the fraction evolves little
with redshift. In contrast, field galaxies from the survey show 
a very strong increase over the same redshift range. It thus appears 
that the environment in the cores of rich clusters is hostile to star
formation at all the redshifts studied. We compare this result with
the evolution of the colours of galaxies in cluster cores, first
reported by Butcher \& Oemler (1984). Using the same galaxies for our
analysis of the [OII] emission, we confirm that the fraction of blue
galaxies, which are defined as galaxies 0.2 mag bluer in the rest
frame $B-V$ than the red sequence of each cluster, increases strongly
with redshift. Since the colours of
galaxies retain a memory of their recent star formation history, while
emission from the [OII] line does not, we suggest that these two results
can best be reconciled if the rate at which the clusters are being
assembled is higher in the past, and the galaxies from which it is
being assembled are typically bluer. 
\end{abstract}

\begin{keywords}
galaxies: clusters: general --- galaxies: evolution --- galaxies: formation
\end{keywords}

\section{Introduction}

More than 20 years ago, Butcher \& Oemler (1978) first reported a
puzzling evolution in the colours of cluster galaxies (the
``Butcher-Oemler effect''). Their results were surprising because they
showed that galaxies in clusters could not have all formed at early
times of the universe and subsequently evolved passively,
which was a popular theory at the time (Eggen, Lynden-bell \& Sandage
1962). The initial results have been confirmed in many subsequent
studies (e.g., Butcher \& Oemler 1984 [BO84]; Couch \& Newell 1984;
Kodama \& Bower 2001; Ellingson et al.\ 2001; Margonnier et al.\
2001). Despite the vintage of these results, there is still
uncertainty over their interpretation. The result remains surprising
since the cluster environment would seem hostile to star
formation. Cluster galaxies undergo efficient ram pressure
(Gunn \& Gott 1972; Abadi, Moore
\& Bower 1999; Fujita \& Nagashima 1999; Quilis, Moore \& Bower 2000),
suffocation due to removal of their
gaseous halos (Larson, Tinsley \& Caldwell 1980; Balogh, Navarro \&
Morris 2000; Drake et al.\ 2000; Diaferio et al.\ 2001;
Okamoto \& Nagashima 2003) and
harassment due to the cumulative effect of fast
encounters with other galaxies (Moore et al.\ 1996, 1999).
Currently, the most popular explanation is that the blue galaxies are
a galaxy population that has arrived only recently in the
cluster. Even though their star formation may be suppressed, their
previous star formation history will still be reflected in their blue
colours (Kauffman 1996; Ellingson et al.\ 2001; for a recent review see
Bower \& Balogh 2004). This scenario also fits well with observations
of a significant population of post-starburst galaxies in clusters
at $z<1$
(Dressler \& Gunn 1992; Postman, Lubin \& Oke 1998; Dressler et al.\ 1999;
Poggianti et al.\ 1999).

On the other hand, the previous discussion does not address why the population
of blue cluster galaxies evolves so strongly with redshift. Although the
rate of infall into clusters does evolve with redshift (Bower 1991;
Kauffmann 1996; Diaferio et al.\ 2001), on its own this effect seems
insufficient to account for the rapid evolution. The additional
component may come from the rapid change in the star formation rates
of field galaxies over the same redshift range (Ellingson et al.\ 2001).
It is then a matter of some debate whether the population of blue galaxies
is a population of galaxies actually in the cluster itself, or whether they
represent a population of infalling galaxies that have not yet reached the
high density region in the cluster core.

Furthermore, colours do not tell the complete story. In order to investigate
the nature of the ``Butcher-Oemler'' effect more directly, we have set out
to measure the star formation rate of galaxies in the cluster
core. H$\alpha$ is the preferred estimator of star formation in the
optical, but a large homogeneous survey covering a wide
range of redshifts has not yet been completed (see Balogh \& Morris
2000; Couch et al.\ 2001; Balogh
et al.\ 2002; Finn, Zaritsky \& McCarthy 2004; Kodama et al.\ 2004 for
some recent results on
distant clusters). The best available alternative is to use the large
and homogeneous sample of clusters that form the CNOC1 survey (e.g.,
Yee, Ellingson \& Carlberg 1996). Although this survey lacks the
wavelength coverage to measure the strength of H$\alpha$, we can
investigate the evolution of star formation activity using the [OII]
$\lambda3727$ emission line. This dataset allows us to investigate the
evolution of the equivalent
width of the [OII] line (EW[OII]) over the redshift range $0.15<z<0.6$.
For a datum at lower redshift, we compare these measurements with
the local 2dF data (Colless et al.\ 2001, 2003). At higher redshifts,
we compare the trends seen in the CNOC1 data with publicly available 
catalogues of four clusters (Cl~1324+3011, Cl~1604+4304, Cl~1604+4321:
Postman, Lubin \& Oke 2001; MS~1054-03: van Dokkum et al.\ 2000).

There has been a
recent realization that the galaxy population is strongly bimodal
(Kauffman et al.\ 2003), with distinct `active' star
forming population and a `passive' population with low levels of
star formation. Within the active population the star formation rate
seems to vary little with environment (Balogh et al.\ 2004b;
Baldry et al.\ 2004) and that it is therefore more useful to
characterise the galaxy population from the ratio of the active and
passive populations. We follow this procedure here, focusing on the
fraction of strongly star forming galaxies (${\rm EW[OII]}>10$\AA) as
a function of redshift. 

This paper is structured as follows. In Section~2, we briefly
describe data used for our analysis.
In Section~3, we investigate the evolution of the fraction of strong
star-forming galaxies at $z\lesssim1$.
We will show that there seems to be little evolution in the star
formation activity within the clusters. Therefore in Section~4, we
compare our results with the evolution of the blue galaxy fraction in
the CNOC1 clusters. We consider the wider implication of these results
in Section~5, and summarise our results in Section~6. Throughout this
paper, we assume a cosmology with $\Omega_0=0.3,$ $\lambda_0=0.7,$
$H_0=70h_{70}^{-1}$~km~s$^{-1}$~Mpc$^{-1}$.

\section{Data}

\subsection{CNOC1}

%
%

\begin{table*}
\caption{A summary of the sample.}
\label{table1}
\begin{tabular}{lccclcccl}
\hline
 & $z$ & $N_{\rm cluster}$ & $N_{\rm field}$
 & \multicolumn{1}{c}{lim. mag}
 & bands$^a$ & \multicolumn{2}{c}{spectroscopy}
 & \multicolumn{1}{c}{Reference} \\
 & & & & & & aperture (kpc)$^b$ & wavelength (\AA) \\
\hline
  2dF    & $0.05<z<0.1$ & 366 & 12426 & $M_B<-19.5$
         & $b_{\rm J}r_{\rm F}$ & 2.0-3.7 & 3600-8000
         & Colless et al.\ (2001,2003) \\
\\
  CNOC1
         & $0.15<z<0.3$ & 180 & 118 & $M_B<-19.5$ & $gr$
         & 3.9-6.7 & 4350-5600 & Yee et al. (1996) \\
         & $0.3<z<0.4$ & 84 & 195 & $M_B<-19.5$
         & & 6.7-8.1 & 4700-6400 \\
         & $0.4<z<0.6$ & 74 & 197 & $M_B<-19.5$
         & & 8.1-10.0 & 4700-7000 \\
\\
  Postman & $0.6<z<0.8$  &  22  & 48  & $M_B<-19.5$ & $BVRI$
          & 6.7-10.5 & 4400-9500 & Postman et al.\ (2001) \\
          & $0.8<z<1$    &  34  & 23  & $M_B<-20.5$ & & 7.5-11.2 \\
\\
  MS1054  & 0.83  &  51  & --  & $M_B<-20$ & F606W,F814W
          & 9.1 & 5700-9500 & van Dokkumn et al.\ (2000) \\
\hline
\end{tabular}
\begin{flushleft}

$^a$ Imaging bands using for deriving the absolute $B$-magnitude in
the rest frame of each galaxy \\
$^b$ derived from the aperture size of the fibre (2dF) or the width of
the slit (others)
\end{flushleft}
\end{table*}

%
%

\begin{figure}
\includegraphics[trim=1mm 1mm 1mm 1mm, clip, width=7cm]{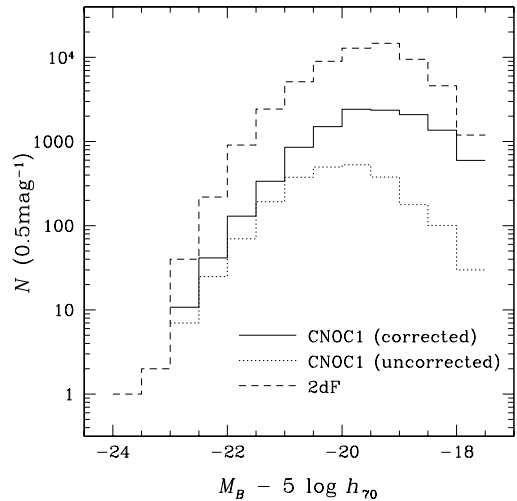}
\caption{The luminosity functions of galaxies in the rest frame $B$
band. The dotted and the solid line
indicate the raw counts and the counts corrected for incompleteness
for CNOC1 galaxies, respectively. The dashed line shows the LF
of 2dF galaxies. Note that these LFs include field galaxies as
well as cluster galaxies.}
\label{fig1}
\end{figure}

The CNOC1 catalogue (e.g., Yee et al.\ 1996)
is a spectroscopic sample of 16 X-ray bright clusters at
intermediate redshift ($0.15<z<0.6$), selected from the EMSS
catalogue (Gioia \& Luppino 1994). Our aim is to study star formation
in high mass systems, thus we exclude E0906, for
which a velocity dispersion could not be computed because of an
apparent double component structure (Carlberg et al.\ 1996).

We calculate the $K$-corrected absolute $B$-magnitude in the rest
frame using the $K$-corrections given in Fukugita, Shimasaku \& Ichikawa
(1995). They are based on Coleman, Wu \& Weedman's (1980) and
Kennicutt's (1992) spectral atlas. The completeness of catalogued
objects is adopted from Balogh et al.\ (1999). We apply a magnitude dependent
weight ($W_m$) to correct for galaxies which are omitted from the 
spectroscopic catalogue.
Figure~1 shows the luminosity function (LF) of CNOC1 galaxies in the
rest-frame $B$ band. The dotted and the solid lines indicate the raw
counts and the counts corrected for spectroscopic incompleteness, 
respectively. We include not only cluster galaxies but also field
ones to check the completeness of the catalogue.
The figure compares the CNOC1 LF
with that of galaxies from the 2dF-galaxy redshift survey (2dF-GRS)
sample (see below). 
It is clear from Figure~1 that incompleteness in the input catalogues
is significant below $M_B>-19.5$ (solid line).
Therefore, we set a limiting magnitude of our analysis as $M_B=-19.5$,
which corresponds to $\sim M^\ast+1$.

Cluster membership is based on the radial velocity
difference from the brightest cluster galaxy, using the method
described in Balogh et al.\ (1999).
We estimate the (projected) radial dependence of the cluster velocity
dispersion, $\sigma(r)$, from the average measured dispersion,
$\sigma_1$, using the mass model of Carlberg, Yee \& Ellingson (1997).
Galaxies with normalized velocities less than 3$\sigma(r)$ are considered
cluster members. We also extract field galaxies from the same CNOC1
area. We treat galaxies with normalized velocities greater than
6$\sigma(r)$ as field galaxies. We use only galaxies within
0.7$h^{-1}_{70}$~Mpc radius, which is a characteristic value of the
cut-off radius adopted by BO84. This facilitates
direct comparison with the Butcher-Oemler effect. Table~1 summarises the
sample we used. Column 1 indicates the data-set; column 2 shows 
the redshift bins into which the data have been split;
column 3 shows the number of cluster galaxies within
0.7$h^{-1}_{70}$~Mpc radius at $M_B<-19.5$; and column 4
indicates the number of field galaxies for which spectroscopy is available.

\subsection{2dF}

We use data from the 2dF galaxy redshift survey (Colless et al.\ 2001,
2003) to set a low redshift datum. We take a volume limited sample of
the data in the redshift range $0.05<z<0.1$, and only galaxies brighter than
$M_B=-19.5$ in order to match the limiting magnitude of the CNOC1
sample (Figure~1).
The 2dF EW[OII] measurement is smoothed with a gaussian kernel of
width 2\AA\ to match the mean error on CNOC1 EW[OII] measurements.

The cluster sample is taken from the redshift space selected 2PIGGZ
cluster and group catalogue (Eke et al.\ 2004). We only use those clusters,
whose velocity dispersion is larger than 500~km/s, and also require
that the number of member galaxies is larger than 30 in order to
exclude poor systems with over estimated velocity dispersion.
Note that even if we adopt more strict conditions for selecting
cluster systems from the 2dF catalogue, the results we show do
not change significantly.
There are 366 galaxies within 0.7$h^{-1}_{70}$~Mpc radius from
the center of clustering systems we adopt, and there are 12426 field
galaxies.

\subsection{Higher redshift data sets}

%
%

\begin{figure}
\includegraphics[trim=1mm 1mm 1mm 1mm, clip, width=7cm]{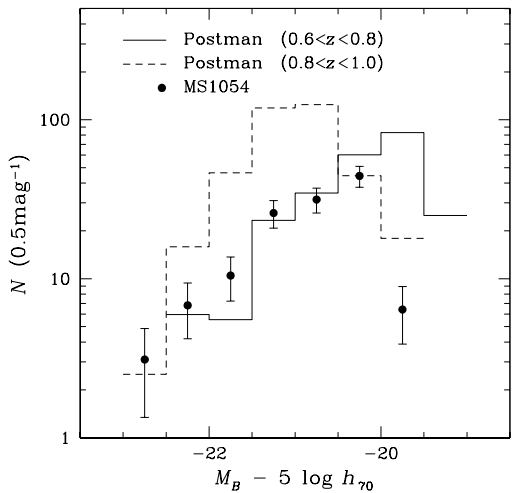}
\caption{The LF of high-$z$ galaxies in the rest frame $B$ band.
The solid and the dashed line indicate the LF of galaxies at
$0.6<z<0.8$ and $0.8<z<1.0$, respectively, which is catalogued
at Postman et al.\ (2001). The filled circles shows the LF of galaxies
in MS~1054. The catalogue of the MS~1054 does not include field
galaxies, while the LFs of Postman's catalogue include galaxies in both
field and cluster environments.}
\label{fig2}
\end{figure}

To investigate the effect of evolution at higher redshift, we added two
publicly available catalogues: Cl~1324+3011 ($z=0.76$), Cl~1604+4304 
($z=0.90$) and Cl~1604+4321 ($z=0.92$)
are from Postman et al.\ (2001) and MS~1054-03 from van Dokkum et al.\
(2000). In addition to the cluster galaxies, the Postman data set 
provides a control sample of high redshift field galaxies.
We refer to these catalogues as the high-$z$ data sets.
Figure~2 shows the LF of galaxies extracted from these catalogue.
The incompleteness of each catalogue is corrected
using literature (Figure~2 of Postman et al.\ 2001 and Figure~1 of van
Dokkum et al.\ 2001). The solid and dashed lines indicate the LF of
galaxies (in both cluster and field environments) at $0.6<z<0.8$
and $0.8<z<1.0$, respectively, extracted from Postman et al.\ (2001).
We find that the incompleteness is small for $M_B<-19.5$
(the limiting magnitude of CNOC1 sample) for galaxies at a
lower redshift bin ($0.6<z<0.8$), while it becomes large at 
$M_B>-20.5$ for galaxies in the higher redshift bin ($0.8<z<1.0$).
Thus, we use only brighter galaxies ($M_B<-20.5$) at $0.8<z<1.0$.
The filled circles of Figure~2 indicate the LF of cluster galaxies
of MS~1054 extracted from van Dokkum et al.\ (2001) (this catalogue
does not include field galaxies). Since the
incompleteness is small to $M_B<-20$ for MS~1054 galaxies,
we use galaxies in the full magnitude range.
We summarise these high-$z$ data-sets at the end of
Table~1. Columns~3 and 4
indicate the number of galaxies at the core ($r<0.7h^{-1}_{70}$~Mpc) of each
cluster and field galaxies, respectively. The limiting magnitude of
the sample is shown in column~5.
We also show the observed imaging bands, which is used for
deriving the $M_B$ of each galaxy in column~6, spectroscopic
apertures derived from the aperture size of the fibre (2dF) or the
width of the slit (others) in column~7, and the spectral range
covered by each observation in column~8.

\section{The evolution of EW[OII] of cluster galaxies}

As we discussed in the introduction, [OII] emission is a workable
tracer of star formation in a statistical sense (e.g., Hopkins et al.\
2003), nevertheless, we will avoid explicit conversion of EW[OII] into
star formation rates. For our purposes, it is sufficient to compare
the line emission properties of high/low redshift and cluster/field
galaxies directly.  This results in a comparison of well defined 
observational quantities.

%
%

\begin{figure}
\includegraphics[trim=1mm 50mm 1mm 60mm, clip, width=8.5cm]{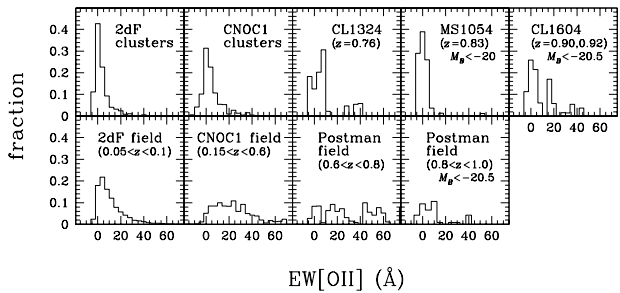}
\caption{Distributions of EW[OII] of the 2dF, CNOC1 and
high-$z$ galaxies. Redshift is increasing from left to right, with
$0.05<z<0.1$ (2dF), $0.15<z<0.6$ (CNOC1) and $0.6<z<1$ (high-$z$
galaxies). The cluster galaxy
sample is shown in the top panels, and the field sample in the bottom
panels, respectively.}
\label{fig3}
\end{figure}

Figure~3 shows the distribution of EW[OII] in 2dF, CNOC1 and
high-$z$ galaxies.
There is a strong spike around ${\rm EW[OII]}=0$\AA\ for cluster galaxies
at all redshifts. This is expected since star formation is known
to be strongly suppressed in galaxy clusters. In contrast, the field
galaxies have a much wider distribution.

As we have discussed, the H$\alpha$ equivalent width distribution
(Balogh et al.\ 2004b; Kauffmann et al.\ 2004) is strongly bimodal.
The bimodality corresponds to a blue, star-forming population and a
red population of galaxies no longer forming stars. The distribution
of EW[OII] is less clearly bimodal due to the less direct and less
sensitive relation between [OII] emission
and star formation. Nevertheless, the distribution can still be usefully
decomposed into two populations by separating the distribution at
a particular EW[OII] value. We have chosen to make this separation at 
${\rm EW[OII]}=10$\AA. The particular choice of
10\AA\ is not crucial, but roughly separates the galaxies with strong
star formation from the passive population.
By examining the objects scattered to negative [OII] values, we can
estimate the level of contamination. With the 10\AA\ cut, less than
1\% of galaxies with intrinsically low star
formation rate scatter into the ${\rm EW[OII]}>10$\AA\ population.

%
%

\begin{figure}
\includegraphics[trim=20mm 1mm 20mm 1mm, clip, width=8cm]{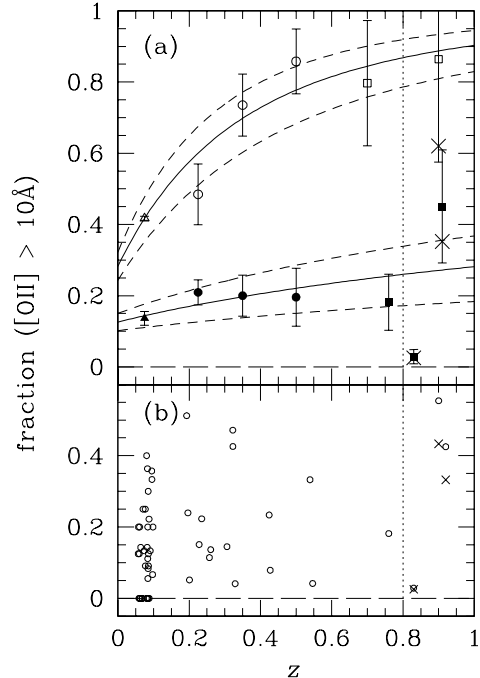}
\caption{The fraction of galaxies with ${\rm EW[OII]}>10$\AA\ as
functions of redshift. (a) The triangles, circles and
squares indicate the 2dF, CNOC1 and high-$z$ galaxies, respectively.
The filled and open symbols correspond to the cluster and the field
galaxy samples, respectively. The error bars indicate the 1$\sigma$
Poisson errors. The two solid lines indicate best fitted lines of
the form $f_{10}=1-a(1+z)^{-b}$ fitted to cluster and field data,
respectively, and the dashed lines show the 1$\sigma$ error of the fits.
(b) The open circle
indicates the fraction of each cluster, which is binned at the panel
(a). The vertical dotted line in the figure separates the data 
sets for where we have had to apply a correction to allow for the
brighter limiting magnitude. The crosses illustrate the fraction
before the correction was made. While the field galaxies show a rapid
evolution in $f_{10}$, the fraction in clusters changes little out
to $z\sim1$.}
\label{fig4}
\end{figure}

In order to investigate the redshift evolution of star formation 
activity, we determine the fraction of star--forming galaxies as those
with ${\rm EW[OII]}>10$\AA\ (hereafter $f_{10}$). 
As Balogh et al.\ (2004b) discusses, the evolution
of this fraction provides a better indicator of the galaxy population
than the median star formation rate. We have confirmed that similar 
trends are seen in the median, but they are more difficult to interpret 
due to the highly skewed distribution of ${\rm EW[OII]}$. 

The evolution of $f_{10}$ is shown in Figure~4(a). 
The triangles, circles and squares indicate the 2dF, CNOC1 and 
high-$z$ data sets,
respectively. The filled symbols correspond to the cluster samples,
while the open symbols correspond to the field galaxy samples.
The limiting magnitudes of the $z>0.8$ data sets do not reach
$M_B=-19.5$ (see \S2.3). The fractions measured for these
systems are shown by the crosses in the diagram. In order to compare
these clusters with the other data points, we must make a correction
for the additional contribution to $f_{10}$ from galaxies below the
magnitude limit. We estimate this by determining the dependence of
$f_{10}$ on limiting magnitude for CNOC1 galaxies at $0.3<z<0.6$
as described below (Figure~5).

As we should expect from previous work on the
cosmic volume averaged star formation rate (Lilly et al.\ 1996;
Madau et al.\ 1996; Madau, Pozzetti \& Dickinson 1998; Hopkins 2004), 
field galaxies show a strong increase in the active galaxy fraction
with redshift. In order to quantify the evolution in $f_{10}$,
we perform least-squares fits using the equation
$f_{10}=1-a(1+z)^{-b}$. For field galaxies, we find $a=0.72\pm0.04$
and $b=2.89\pm0.74$.

In contrast, the fraction of active galaxies is
always low in the cluster galaxies, and there is little discernible increase 
with redshift as far as $z\sim0.8$. In order to perform
least-squares fits, we combine the distant clusters at $z>0.6$
(Cl~1324, MS~1054, Cl~1604+4302, and Cl~1604+4321) from the different
surveys into a single bin. We find that $a=0.87\pm0.02$ and
$b=0.28\pm0.14$ for cluster galaxies. Thus, the parameter $b$, which
indicates an evolutionary strength, of cluster galaxies is inconsistent
with that of field galaxies with $\sim3\sigma$ level.
This result is in agreement
with the weak evolution in the cluster core ``emission'' population 
(derived from a principle component analysis) reported for the CNOC1 alone 
clusters by Ellingson et al.\ (2001), but extends the trend to both lower 
and higher redshift.  We will contrast this result with the evolution
of the blue fraction in clusters reported by BO84
in the following section.

Figure~4(b) shows the fraction of star forming galaxies of each individual
cluster. The variation in $f_{10}$ is large at all redshifts due to
the small numbers of galaxies in each individual cluster. It is 
possible to find individual clusters with $f_{10}$ as low as 0, or 
as high as 0.6. From the limited data that is currently available,
the  scatter appears similar at both low and high redshift.
The apparent inconsistency between MS1054 (which has a very low $f_{10}$)
and the CL1604 clusters (which have significantly higher values) appears 
entirely consistent with the variations seen in lower redshift clusters.
It is important to realise, however, that the highest redshift sample is 
limited to only 4 clusters, and thus has limited sampling of the range
of $f_{10}$ in clusters at this redshift. 

The fraction of emission--line or blue galaxies within a given
projected radius is an overestimate of the blue fraction within the
corresponding physical radius, because of the strong colour gradient in
clusters. We can estimate the effect of projection from the simulations
of Diaferio et al.\ (2001; their Figure~13).  This work shows that within
a projected radius of $\sim 0.7h^{-1}_{70}$~Mpc, the
fraction of interlopers (defined as galaxies with physical distances
beyond $R_{200}$) depends strongly on galaxy colour: $\sim 60$ per cent
for blue galaxies and only $\sim 10$ per cent for red galaxies. Making
a correction for these interlopers reduces our blue (or star--forming)
fractions in clusters from $\sim 20$ per cent to $\sim 10$ per cent.
Moreover, if the fraction of blue interlopers increases with increasing
redshift (as expected due to the colour evolution in the field), this
will make the observed trend of $f_{10}$ with redshift 
in clusters {\it even weaker}. Thus accounting for projection effects
only strengthens our conclusions.

%
%

\begin{figure}
\includegraphics[trim=1mm 40mm 1mm 40mm, clip, width=8cm]{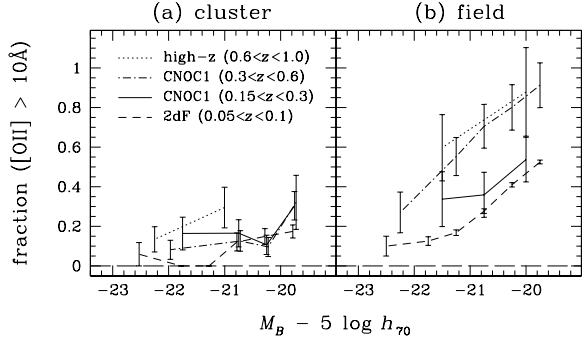}
\caption{The fractions of galaxies with ${\rm EW[OII]}>10$\AA\ in the
(a) cluster and (b) field as functions of the rest-frame $
B$ magnitude. The solid and dot-dashed lines indicate the CNOC1
galaxies at $0.15<z<0.3$ and $0.3<z<0.6$, respectively. The dashed and
dotted lines show the 2dF and high-$z$ galaxies, respectively.}
\label{fig5}
\end{figure}

In order to investigate the dependence on limiting magnitude in the
data-sets we have used, we determined
$f_{10}$ for different luminosity bins. The result is shown in
Figure~5. The solid and dot-dashed lines indicate the CNOC1 galaxies
at $0.15<z<0.3$ and $0.3<z<0.6$, respectively, and the dashed lines
show 2dF galaxies. We also show a
result of high-$z$ galaxies with the dotted lines. Because of the low numbers
of galaxies and variable magnitude limits we only plot two points 
for high-$z$ galaxies.

It is clear that the fainter galaxies in the field are more
likely to be star forming and possess larger EW[OII] than brighter
ones (Kauffmann et al.\ 2003; Baldry et al.\ 2004; Balogh et al.\ 
2004a). On the other hand, the trend is much weaker
for galaxies in cluster cores over the magnitude range probed by this
data. The plot also confirms the trends seen in Figure~4:
the fraction of star forming galaxies is always greater in the field than 
in the cluster and that the redshift evolution is seen for field 
galaxies, but not for cluster galaxies.

\section{Comparison with blue galaxy fractions}

In the previous section we find that the EW[OII] of galaxies in
clusters shows little evolution at $z<1$. At face value, this result 
seems to be inconsistent with the Butcher-Oemler effect: the
increase in the fraction of blue galaxies in cluster cores with redshift
(Butcher \& Oemler 1978; BO84; Couch \& Newell 1984).
The difference cannot be due to the cluster sample used by
Kodama \& Bower (2001) and Ellingson et al.\ (2001) both found
a significant Butcher-Oemler 
effect using the CNOC1 clusters. However, among the sample used
by Kodama \& Bower (2001), only bright galaxies are spectroscopically
confirmed requiring them to apply a foreground/background subtraction 
using a general field data for faint galaxies. This enabled them to 
reach $M_V<-20+5\log~h_{50}$, consistent with the original definition 
of the limiting magnitude of Butcher-Oemler effect.
Ellingson et al.\ (2001) investigate blue fractions of CNOC1
galaxies using spectroscopically confirmed members, however, the radius 
constraint they adopted is
a dynamically determined value, $R_{200}$, the radius within which
the average cluster density is 200 times the critical density (e.g.,
Carlberg et al.\ 1996, 1997). The
$R_{200}$ of CNOC1 clusters is 1.07-2.24~$h^{-1}_{70}$~Mpc, which is
wider than our
radius constraints ($0.7h^{-1}_{70}$~Mpc). In order to eliminate
the possibility that the increasing trend may be
due to galaxies in the outer cluster region, we
directly compare with measurements of the evolution in EW[OII], by
recomputing the blue fraction only from the
spectroscopically confirmed galaxies and adopting the same
limiting magnitude and radial cut-off.

We use the $g-r$ colour in Thuan-Gunn system (Thuan \& Gunn 1976),
which is available in the CNOC1 catalogue to discriminate between blue
and red galaxies. On the $g-r$ versus $r$ diagram we fit the
colour-magnitude relation for the early-type galaxies using a biweight
fit (Beers, Flynn \& Gebhardt 1990). Blue galaxies are defined as
galaxies 0.2 mag bluer in the rest frame $B-V$ than the red-sequence
of each cluster. We calculate the difference in $g-r$ colour in the
observed bands corresponding to $\Delta(B-V)=-0.2$ in the rest frame
(BO84; Kodama \& Bower 2001).
This corresponds to $\Delta(g-r)$ of 0.20-0.44 mag, depending on the
cluster redshift.
As discussed above, we fix the radial cut off at
0.7$h_{70}^{-1}$~Mpc (this is typical of the $R_{30}$ cut off used by BO84).
Note that our magnitude limit is $M_B=-19.5$, which is brighter than
the limit of BO84, $M_V=-19.3$.

%
%

\begin{figure}
\includegraphics[trim=1mm 1mm 1mm 1mm, clip, height=7cm]{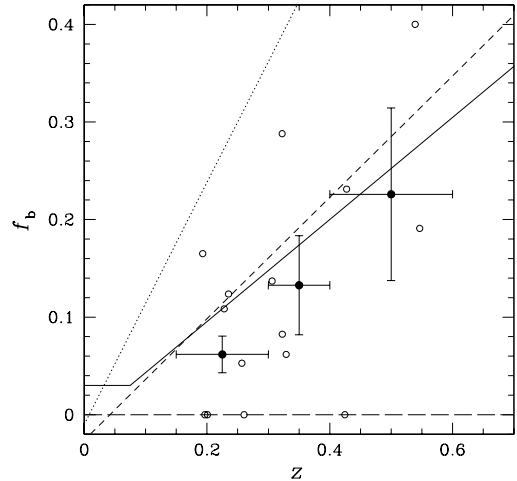}
\caption{The blue galaxy fraction $f_{\rm b}$ for CNOC1 clusters
as a function of
redshift. The filled circles show the average blue fraction in each
redshift bin. The small open circle indicates the fraction for each cluster.
The solid line are extrapolations to higher redshifts of
the observed evolution of $f_{\rm b}$ obtained by BO84.
The dashed and dotted line indicate the $f_{\rm b}$ derived by
Ellingson et al.\ (2001) and Margonnier et al.\ (2001), respectively.
}
\label{fig6}
\end{figure}

Figure~6 shows the calculated blue fraction as a function of redshift.
The filled circles
show the average blue fraction from the CNOC1 data. The small open circles 
show the blue fraction of each individual cluster.
The solid line in Figure~6 is a linear
extrapolation to higher redshifts of the observed evolution of $f_{\rm
b}$ obtained by BO84, while 
the dashed line and the dotted line indicate the best-fit lines from
Ellingson et al.\ (2001) and Margonnier et al.\ (2001), respectively.  
In contrast to the results we obtained from our analysis of EW[OII],
we see that the blue galaxy fraction of CNOC1 clusters has an
increasing trend with redshift. Although the uncertainty is large, the
slope of this relation is  consistent with BO84, Kodama \& Bower
(2001) and Ellingson et al.\ (2001).
Note that even if we make the colour boundary redder by 0.1 mag
($\Delta(B-V)=-0.1$), the increasing trend of the $f_{\rm b}$ does not
change significantly.
As the magnitude limits are different between our analysis and BO84,
it is not surprising that our results fall slightly below the trend proposed by
BO84. This reflects the trend for a larger fraction of fainter galaxies
to have blue colours (Kodama \& Bower 2001; Kauffmann et al.\ 2003;
See also Figure~5). 
Our $f_{\rm b}$ values are also slightly lower than those of 
Ellingson et al.\ (2001), a difference that arises from the
different radial constraints we have applied: we use only galaxies at
the cluster
core ($r<0.7h_{70}^{-1}$~Mpc), while their sample include galaxies out
to $R_{200}$. The extremely rapid change in $f_{\rm b}$ over the
redshift range out to $z\sim0.2$ reported by Margonnier et al.\ (2001)
is not consistent with our study.

\section{Discussion}

%
%

\begin{figure*}
\includegraphics[trim=1mm 55mm 1mm 55mm, clip, width=18cm]{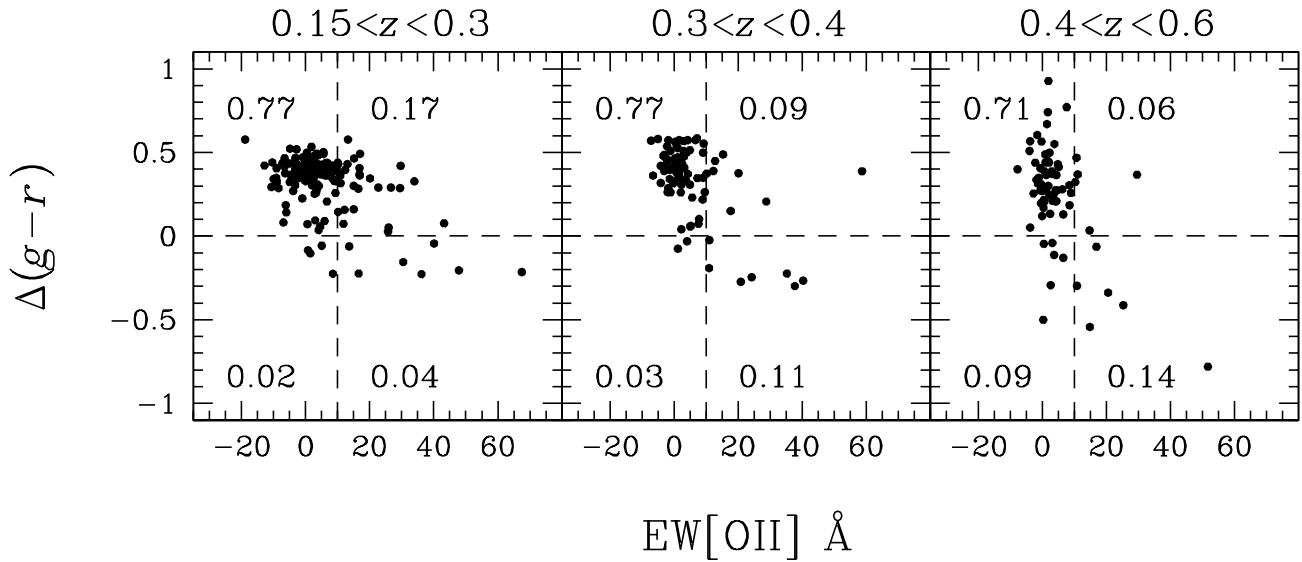}
\caption{EW[OII] versus $\Delta(g-r)$ diagrams of the CNOC1 cluster 
galaxies. The vertical axis shows the `differential colour', defined
as the observed colour minus the dividing colour between blue and
red galaxy (see text for detail) at the same $M_B$ magnitude. The
sequence of panels shows 
clusters at increasing redshift. At higher redshift, the blue galaxy
population becomes increasingly offset from the red colour sequence.
We also show the fraction of galaxies in each of the
four regions.}
\label{fig7}
\end{figure*}

We set out to investigate the evolution of the [OII] line in cluster
galaxies. We found that the fraction of cluster galaxies with
strong [OII] emission was always small and evolved little with
redshift. Viewed in isolation, this result is not surprising: 
the very dense environment
in cluster cores is hostile to star formation. The commonly discussed
environmental processes of ram-pressure (Gunn \& Gott 1972; Abadi et
al.\ 1999; Fujita \& Nagashima 1999; Quilis et al.\ 2000), harassment
(Moore et al.\ 1996, 1999) and strangulation (Larson et al.\ 1980;
Balogh et al.\ 2000; Drake et al.\ 2000; Diaferio et al.\ 2001; Okamoto \&
Nagashima 2003) are all effective in this region. Our results are showing
us that this environment has always been hostile even out to $z\sim1$.
The very dense environment of the cluster core becomes increasingly
rare as we look back to higher redshift, but star formation is always
highly suppressed there.

What is puzzling is to compare this result to the evolution in the
colours of cluster galaxies. A long history of studies have reported
an increase in the fraction of blue galaxies in cluster cores at higher
redshifts (BO84; Kodama \& Bower 2001;
Ellingson et al.\ 2001; Margonnier et al.\ 2001). With the CNOC1 data
set, we are able to compare the two measurements of galaxy star
formation activity directly. Within the large uncertainties, the
colours of the galaxies confirm the increase in blue fraction
suggested by BO84.

First we consider two biases in the measurements that could affect
this result. (1) Aperture effects are of course a worry in assessing
these results. However, Appendix A of Wilman et al.\ (2004) makes a
detailed study of the likely
effect on [OII] emission by examining the radial dependence of the
colours of galaxies in the Sloan Digital Sky Survey (hereafter SDSS;
York et al.\ 2000; Stoughton et al.\ 2002; Abazajian et al.\ 2003). He
concludes that there is
currently no evidence that the fraction of [OII] emitting galaxies
determined from relatively small apertures (eg., by the 2dF and SDSS
fiber spectrographs) is substantially different from that determined
through larger apertures.
(2) If the measurement error on the [OII] emission was large, a weak trend
in the fraction of [OII] emitters could be masked by objects with intrinsically
low star formation rates being scattered into the ${\rm
EW[OII]}>10$\AA\ region.
However, the estimated errors in the CNOC1 equivalent
widths are only 2\AA\ on average, and thus the error is too small
to account for the tail seen in Figure~3 as we have shown in \S3.

If we reject both of these measurement biases,
how can we reconcile these apparently contradictory results? The key
may lie in the way the two measures of the galaxy activity are related to the
star formation history.  While the [OII] emission measures the radiation
from short-lived hot stars, the $g-r$ colour contains a memory of the
star formation rate over the previous few Gyr (eg., van Dokkum et al.\
2000). Thus one possible explanation
for the result is that while most galaxies in these central cluster
regions currently have little star formation at all redshifts, more
galaxies in the higher redshift clusters have experienced significant
star formation within the last few Gyr, and have evolved from a bluer
starting point. In Figure~7, we directly compare the EW[OII] with
the offset of the galaxy from the dividing colour between blue and red
galaxy, which is defined at \S4
($\Delta(g-r))$.  At higher redshift, there is a greater spread in 
colour for the range of EW[OII] (due both to the $K$-correction, and
the greater star formation rates of field galaxies) while the range
in EW[OII] is roughly similar.  Thus the star forming population in the lower 
redshift bin lie closer to the colour threshold than to the EW[OII]
threshold,  while the converse is true in the high redshift panel.
This provides some support for the differential evolution of EW[OII]
and colour that we have discussed.

If this is a satisfactory explanation for the different evolution of 
$f_{10}$ and $f_b$, can we understand what drives it? Two possibilities
stand out: (1) the more rapid assembly of clusters at higher redshift
(Bower 1991; Kauffmann 1996) or (2) the star formation rates in the
galaxies that are being accreted by the local clusters are lower than
in their high redshift counter parts. In the
first case, the mass assembly history of clusters differs between
low and high redshift because clusters are much rarer objects at $z\sim0.5$
than at the present-day, and because the cosmic time available
to build the cluster is less at higher redshift. 
In the second case, the reduction in the
star forming fraction of the infalling galaxies is clearly shown
by our data on the field galaxies. As a result the galaxies
which are accreted by the local clusters are much less likely to have
significant star formation in the first place. It is likely that both
effects contribute, and hard to disentangle them.

An alternative possibility is that the [OII] emission is powered by AGN
activity rather than by star formation. In this case, the fractions
would reflect very different physical processes, whose evolution might very 
well differ. The scenario naturally explains the existence of red galaxies
with detectable [OII] seen in the first panel of Figure~7. 
Martini et al. (2004) show from X-ray observations that clusters may
contain a significant population of AGN. Often these galaxies cannot be 
recognised
as AGN from their optical spectra alone, yet they are clearly luminous
point sources in the X-ray data. Their results suggest that the active
systems are over represented in clusters compared to the field. It is 
thus possible that an underlying trend in the star formation rates of
cluster galaxies is masked by a opposing trend in the fraction undergoing
AGN activity. It should be noted, however, that the average AGN fraction
is significantly less that 5\% in their data.

In future, it will also be possible to carry out this programme using
other indicators of star formation and to investigate other galaxy
environments. Wilman et al.\ (2004) have used [OII] to investigate the
evolution of galaxies in groups. Interestingly, they find that the
fraction of star forming galaxies in the group environment increases
significantly over the redshift range $z=0$ to 0.5, suggesting that
group galaxy properties are much more closely tied to galaxies in the
field than galaxies in cluster cores. While the [OII] line is conveniently
placed in the optical spectral window, it suffers the disadvantage
of being sensitive to dust and metallicity as well as the star formation
rate. The H$\alpha$ line gives a cleaner measure of the star formation rate.
For example, Finn et al.\ (2004) report measurements of H$\alpha$ emission 
from a poor cluster (Cl~J0023+0423B) at $z=0.845$. Although their measure of
activity is not directly comparable to the one we use here, they find a 
much higher incidence of star formation activity than any of our clusters,
or that of our H$\alpha$ measurements in the rich cluster Cl0024+1652 
(Kodama et al.\ 2004). At present, however, the sample of clusters
with well sampled H$\alpha$ determinations is too small to make an 
accurate test of the rate of evolution. Furthermore,
it is harder to measure over a wide range of redshifts because it is
redshifted out of the optical window. Nevertheless, new multi-object
spectrographs working in the near infrared, such as MOIRICs, will allow
us to probe the star formation rates in distant clusters with much
greater accuracy using H$\alpha$ emission. Alternatively, it will be
possible to use the Spitzer satellite to measure the star formation
rates in clusters on the basis of their far infra-red dust emission
(Duc et al.\ 2002; Stanford et al.\ in prep). 
The prospects look extremely good for extending the work we have described 
here.

\section{Conclusions}

By compiling a homogeneous sample of the clusters with extensive 
spectroscopic measurements of the [OII] emission line,
we have presented an analysis of the evolution of the fraction 
of strongly star forming galaxies in cluster cores as a function of 
redshift.  Our sample is based on the CNOC1 cluster survey (Yee et al.\ 1996), 
but extends this data set to lower and higher redshift using the 2dF-GRS 
(Colless et al.\ 2003) and spectroscopic surveys of higher
redshift clusters by Postman et al.\ (2001) and van Dokkum et al.\ (2000).

Our results show little evolution in the fraction of galaxies
with EW[OII] $>$ 10\AA. We constrain the rate of evolution to be
$\Delta f_{10}/\Delta z<0.2$. This result suggests that the 
cluster environment has always been hostile to star formation, at 
least for $z<1$. Many mechanisms have been proposed to explain the 
hostility of the cluster environment at low redshift. Our results show
that the same mechanisms also operate effectively in this special
environment at higher redshift, despite the higher rates of galaxy
infall that are expected (Kaufman 1996) and the higher fraction of 
star forming galaxies in the field population.
It is important to note, however, that there is extensive scatter
between individual clusters at all redshifts; thus individual clusters
may pass through episodes with higher rates of field galaxy accretion.

The slow evolution of $f_{10}$ we report, initially seems at odds with
the strongly increasing fraction of blue galaxies in clusters reported 
in previous papers (eg.\ BO84); however, we confirm that the fraction
of blue galaxies increases strongly with redshift using the same
galaxies for our analysis of the [OII] emission.
We consider how the results can be reconciled.
Since the [OII] emission line is an instantaneous measure of the star 
formation rate, while galaxy colour reflects the average star formation 
rate over the previous few Gyr,
the two observations may be compatible if the rate at which clusters
are assembled is sufficiently high at high redshift and the galaxies 
from which the system is assembled are initially much bluer.  Another 
possibility is that the [OII] emission line galaxies that we see in the
low redshift systems may be powered by AGN activity rather than by 
AGN activity.  A larger compilation of deep X-ray data is required
to exclude this possibility.

The results we present are limited by the small numbers of clusters which
have been extensively surveyed at high redshift, and by the limitations
of the [OII] emission line as a star formation rate indicator. The prospects
for extending this work using H$\alpha$ line strengths, SPITZER mid-IR
fluxes and GALEX ultra-violet photometry as star formation rate indicators 
look very encouraging.

\section*{Acknowledgments}

Many thanks to Erica Ellingson and the CNOC1 collaboration, for
providing us with the full CNOC1 catalogue for this analysis. We thank
Ian Lewis for providing us with the [OII] measurement from 2dF-GRS
prior to their publication. This
project has made extensive use of the NASA Extragalactic Database (NED)
operated by the Jet Propulsion Laboratory, Caltech.
RGB is supported by a PPARC Senior Research Fellowship. FN, MLB and
DJW also acknowledge the financial support from UK PPARC.  We thank
the anonymous referee for their helpful suggestions.

\label{lastpage}

\end{document}